\documentstyle[12pt]{report}
\def\thebibliography#1{\leftline{\large \it References}\list
  {[\arabic{enumi}]}{\settowidth\labelwidth{[#1]}\leftmargin\labelwidth
    \advance\leftmargin\labelsep
    \usecounter{enumi}}
    \def\newblock{\hskip .11em plus .33em minus .07em}
    \sloppy\clubpenalty4000\widowpenalty4000}

\def\nn{\nonumber}
\def\ee{\end{eqnarray}}
\def\be{\begin{eqnarray}}

\def\Tr{\mbox{Tr}}
\def\skp{\epsilon}

\renewcommand{\thefootnote}{\fnsymbol{footnote}}
\begin{document}
\begin{titlepage}

\begin{center}

\rightline{UNITU--THEP--21/1996}
\rightline{TAN-FNT-96/007}
\rightline{October 1996}

\vskip 2.0cm

{\Large\bf Strangeness Dependence in \\ \vskip0.5cm
Radiative Hyperon Decay Amplitudes\footnote[1]{Supported in 
part by the Deutsche Forschungsgemeinschaft (DFG) under contract 
Re 856/2--2.}}
\vskip 1.5cm

{T. Haberichter$^a$, H. Reinhardt$^a$, 
N. N. Scoccola$^b$\footnote[2]{Fellow
of the CONICET, Argentina.} and H. Weigel$^a$}
\vskip 1cm
{\it
$^a$ Institute for Theoretical Physics, T\"ubingen University,\\
     Auf der Morgenstelle 14, D--72076 T\"ubingen, Germany.\\
$^b$ Physics Department, Comisi\'on Nacional de  
     Energ\'{\i}a At\'omica,\\
     Av.Libertador 8250, (1429) Buenos Aires, Argentina. }

\vskip 2.cm
{\bf ABSTRACT}\\
\begin{quotation}
The radiative decays of the $\frac{3}{2}^+$ baryons are studied 
in the three flavor generalization of the Skyrme model. The kaon 
fields are treated in the slow rotator approach which properly 
accounts for the observed deviations from the $U$--spin relations 
for the hyperon magnetic moments. This makes possible a critical 
discussion of the $U$--spin selection rules for the radiative 
hyperon decays. The variation of the decay widths with strangeness 
is studied and a comparison with other treatments of the $SU(3)$ 
Skyrme model is performed in order to analyze the effects of 
flavor symmetry breaking.
\end{quotation}
\end{center}
\vskip 0.5cm
\leftline{\it PACS: 11.10.Lm, 12.39.Dc, 13.40.Hq.}
\vskip 0.5cm
\leftline{\it Keywords: \parbox[t]{15cm}{Skyrmion, Flavor 
Symmetry Breaking, Radiative Hyperon Decays,\\ $U$--Spin Relations.}}

\end{titlepage}

\setcounter{page}{2}
\renewcommand{\thefootnote}{\arabic{footnote}}

\leftline{\large \it 1. Introduction}
\stepcounter{chapter}
\bigskip

At present, only few data are available on the electromagnetic
decays of the $\frac{3}{2}^+$ baryons. Although recently the 
reaction $\Delta\to N\gamma$ has carefully been analyzed at 
MAMI \cite{Be95} the decay parameters are still unknown for those 
$J=\frac{3}{2}$ to $J=\frac{1}{2}$ transitions, which involve strange 
baryons.  Upcoming experiments at CEBAF \cite{CEBAF} and Fermilab
\cite{Fermilab} are expected to provide some data on these
radiative decays soon and thus give more insight in the 
pattern of flavor symmetry breaking. Exhaustive studies of the 
radiative hyperon decays have been performed in a number of models 
which include the non--relativistic quark model \cite{Da83,Le93}, 
the MIT bag model \cite{Ka85}, heavy baryon chiral perturbation 
theory \cite{Bu93}, a quenched lattice calculation \cite{Le93} as 
well as recent Skyrme model studies \cite{Sch95,Ab96}.

A particular feature is that $U$--spin symmetry would imply that 
the transition matrix elements vanish for the processes which 
involve the negatively charged hyperons, {\it i.e.} 
$\Sigma^{*-}\rightarrow\gamma\Sigma^-$ and 
$\Xi^{*-}\rightarrow\gamma\Xi^-$ \cite{Li73}. Although $U$--spin 
symmetry is not exactly realized in nature this result has recently 
been verified approximately \cite{Sch95,Ab96} in both the bound state 
(BSA \cite{Ca85}) and rigid rotor (RRA \cite{Gu84}) approaches to 
the $SU(3)$ Skyrme model \cite{Sk61,Ad83,Ho86,We96}. In the Skyrme model 
baryons emerge as solitons configurations of the pseudoscalar 
mesons. These two approaches conceptually differ in the way the 
kaon fields are treated. The RRA starts from a flavor symmetric 
formulation wherein non--vanishing kaon fields arise from a rigid 
rotation of the classical pion field. The associated collective 
coordinates, which parametrize these large amplitude fluctuations 
off the soliton, are canonically quantized to generate states which 
possess the quantum numbers of physical hyperons. It turns out that 
the resulting collective Hamiltonian can be diagonalized exactly even 
in the presence of flavor symmetry breaking \cite{Ya88}. As a result 
the baryon wave--functions significantly deviate from those obtained 
in the flavor symmetric version of the model. On the other hand the 
BSA treats the kaons as small amplitude fluctuations off the soliton,
{\it i.e.} a flavor symmetric formulation is completely waived. 
As a consequence of flavor symmetry breaking a bound state with unit 
strangeness charge develops out of the zero mode \cite{Ca85}. 
Hyperons are then constructed by pertinent occupations of this bound 
state while the $SU(2)$ quantum numbers are generated analogously to 
the two flavor Skyrme model \cite{Ad83} from the large amplitude 
fluctuations in coordinate-- and iso--space. Despite that in both 
the RRA and BSA the baryon wave--functions significantly deviate from 
the $SU(3)$ symmetric ones it is nevertheless not surprising that the 
$U$--spin selection rules for the radiative hyperon decays are almost 
preserved. The reason is that from studying the magnetic moments of 
the $\frac{1}{2}^+$ baryons it is known that the experimentally 
observed $U$--spin violation requires a strangeness dependent 
classical meson configuration \cite{Sch92,Sch91}. Such a dependence is 
not incorporated in these two approaches. A similar statement can be 
made for the hyperon radii, for which also a sophisticated treatment 
of symmetry breaking is required to reproduce the empirical pattern 
\cite{Po90}. The main purpose of the present study is to address the 
question of strangeness dependence of the decay widths for radiative 
hyperon decays. Of course, this also includes the role of the 
$U$--spin selection rule for the processes 
$\Sigma^{*-}\rightarrow\gamma\Sigma^-$ and
$\Xi^{*-}\rightarrow\gamma\Xi^-$. As argued, a treatment should be
employed, which is capable of reproducing the pattern of flavor 
symmetry breaking of other electromagnetic observables. For 
definiteness we will employ the slow rotator approach (SRA) to the 
Skyrme model which has been shown to reproduce the experimental pattern 
of both the hyperon magnetic moments and radii \cite{Sch92}. The 
starting point for the SRA essentially is the RRA, however, the 
stationary equation is solved for each orientation in flavor space. 
This procedure yields a strangeness dependent soliton. The underlying 
picture is that the collective rotation of the soliton proceeds slowly 
enough that the profile function can react according to the forces 
exerted by flavor symmetry breaking; whence the notion slow rotator.

The paper is organized as follows: In section 2 we will briefly review 
the appearance of electromagnetic fields in the Skyrme model as well 
as the SRA to the three flavor version of the model. Subsequently we 
will discuss the pertinent matrix elements in section 3 and then 
present the numerical results for the radiative hyperon decay widths 
in section 4. We also compare our predictions with other treatments 
of the Skyrme model as well as available predictions of other models. 
Concluding remarks may be found in section 5. Some technicalities 
are relegated to a short appendix.

\bigskip
\leftline{\large \it 2. The Model}
\stepcounter{chapter}
\bigskip

Our starting point is a gauged effective chiral action with 
appropriate symmetry breaking terms. In the case of three flavors 
it is a functional of the pseudoscalar octet $\phi$ and the photon 
field $A_\mu$, the former is non--linearly represented by the chiral 
field $U={\rm exp}(i\phi)$. For a convenient presentation we split the 
action into four pieces
\be
\Gamma = \Gamma_{SK} + \Gamma_{an} 
+ \Gamma_{sb} + \Gamma_{non-min} \ .
\label{lag}
\ee
The first term represents the gauged Skyrme action
\be
\Gamma_{SK} =
\int d^4 x \Big\{ {f^2_\pi \over 4} 
\Tr\left[ D_\mu U (D^\mu U)^\dagger \right]
+
 {1\over{32 \skp^2}}
 \Tr\left[ [U^\dagger D_\mu U , U^\dagger D_\nu U]^2\right] \Big\} \, .
\ee
Here $f_\pi=93{\rm MeV}$ is the pion decay constant and $\epsilon$ is 
the dimensionless Skyrme parameter. The covariant derivative is defined 
via the electric charge matrix ${\cal Q}$
\be
D_\mu U=\partial_\mu U + ie\ A_\mu \ [{\cal Q}, U ] \ ,
\qquad 
{\cal Q}=\frac{1}{2}\ \left[\lambda_3+ \frac{1}{\sqrt3}\lambda_8
\right] \, ,
\ee
Henceforth we adopt Gaussian units, {\it i.e.} $e^2=1/137$. 
$\Gamma_{an}$ is the Wess-Zumino action gauged to contain the photon 
field \cite{Wit83}:
\be
\Gamma_{an} &=& - \frac{i N_C}{240 \pi^2}
\int d^5x \ \epsilon^{\mu\nu\rho\sigma\tau}\
\Tr[ L_\mu L_\nu L_\rho L_\sigma L_\tau] \nn \\
& & - \frac{N_C}{48\pi^2}
\int d^4 x \ \epsilon^{\mu\nu\rho\sigma} 
e A_\mu \ \Tr\left[{\cal Q} \left( L_\nu L_\rho L_\sigma -
     R_\nu R_\rho R_\sigma \right) \right] \ + \ {\cal O}(e^2 A^2).
\label{WZ}
\ee
Here we have used  $L_\mu = U^\dagger \partial_\mu U$ and
$R_\mu = U \partial_\mu U^\dagger$. Furthermore 
$N_C=3$ is the number of colors. The flavor symmetry 
breaking terms are contained in $\Gamma_{\rm sb}$ 
\be
\Gamma_{sb} & = &\int d^4x \left\{
 { f_\pi^2 m_\pi^2 + 2 f_K^2 m_K^2 \over{12} }
 \Tr \left[ U + U^\dagger - 2 \right] 
+{f_\pi^2 m_\pi^2 - f_K^2 m_K^2 \over{2\sqrt{3}}}
\Tr \left[ \lambda_8 \left( U + U^\dagger \right) \right] \right.
\nonumber \\
& & \qquad
 \left.
+ { f_K^2 - f_\pi^2\over{4} }
\Tr \left[ {\hat S} \left(
U (D_\mu U)^\dagger D^\mu U + 
U^\dagger D_\mu U (D^\mu U)^\dagger \right)
\right] \right\} \, ,
\label{sb}
\ee
\noindent
where ${\hat S}={\rm diag}(0,0,1)$ is the strangeness projector.
In eq (\ref{sb}) $f_K$ is the kaon decay constant while $m_\pi$ and 
$m_K$ are the pion and kaon masses, respectively. At non--vanishing 
momentum transfer non--minimal couplings of the photons to the 
pseudoscalar fields may be relevant
\be
\Gamma_{non-min}=i\int d^4x
L_9\left(\partial_\mu A_\nu -\partial_\nu A_\mu\right)
{\rm Tr}\left[{\cal Q}\left(L^\mu L^\nu+R^\mu R^\nu\right) 
\right] \ .
\label{lagl9}
\ee
In fourth order chiral perturbation this term is needed to correctly 
reproduce the electromagnetic pion radius thereby determining the 
dimensionless coefficient $L_9=(6.9\pm0.7)\times10^{-3}$ \cite{GL}.
The full action (\ref{lag}) defines the electromagnetic current 
$J_\mu$ via the expansion in the photon field
\be
\Gamma[U,A_\mu]=\Gamma_{strong}[U]+e\int d^4x\ J_\mu A^\mu 
+{\cal O}(e^2 A^2)\ .
\label{defcur}
\ee
The resulting covariant expression for $J_\mu$ in terms of the 
chiral field $U$ may readily be taken from the literature
\cite{Pa91,Go96}:
\be
J_\mu&=&-\frac{f_\pi^2}{2}\Tr\left\{{\cal Q}
\left(L_\mu+R_\mu\right)\right\}
+\frac{1}{8\epsilon^2}\Tr\left\{{\cal Q}\left(
[L_\nu,[L_\mu,L^\nu]]+[R_\nu,[R_\mu,R^\nu]]\right)\right\}
\nonumber \\ && 
-\frac{N_C}{48\pi^2}\epsilon_{\mu\nu\rho\sigma}\Tr\left\{{\cal Q}
\left(L^\nu L^\rho L^\sigma-R^\nu R^\rho R^\sigma\right)\right\}
-iL_9\Tr\left\{{\cal Q}\partial^\nu\left(
[L_\nu,L_\mu]+[R_\nu,R_\mu]\right)\right\}
\nonumber \\ &&
-\frac{f_K^2-f_\pi^2}{4}\Tr\left\{{\cal Q}\left(
\{U{\hat S}+{\hat S}U^\dagger , L_\mu\}+
\{{\hat S}U+U^\dagger {\hat S} , R_\mu\}\right)\right\} .
\label{curr}
\ee

In order to generate baryon states of good spin and flavor quantum 
numbers we consider the solitonic meson configuration that corresponds 
to an arbitrary orientation of the hedgehog $U_0$ in flavor space,
{\it i.e.} $U=AU_0A^\dagger$. For the time being we confine ourselves to
static rotations in order to establish the slow rotator approach. 
Due to spin and isospin invariance this corresponds to the 
{\it ansatz} \cite{Sch92}
\be
U(\mbox{\boldmath $r$},\nu)=
{\rm exp}\left(-i\nu\lambda_4\right)\
{\rm exp}\Big(i{\mbox{\boldmath $\tau$}}\cdot
\hat{\mbox{\boldmath $r$}}F(r,\nu)\Big)\
{\rm exp}\left(i\nu\lambda_4\right) 
\label{slowrot}
\ee
for the chiral field. As already indicated this {\it ansatz} 
has the remarkable feature that the chiral angle may depend on the 
flavor orientation, which is characterized by the strangeness changing 
angle $\nu\in[0,\pi/2]$. Substituting this {\it ansatz} into the 
strong interaction part of the action, $\Gamma_{strong}$, yields the 
classical energy $E(\nu,F]$ as a functional of the chiral angle $F$ 
and a function of the strangeness changing angle $\nu$. It 
is important to note that the explicit dependence on $\nu$ originates 
from the symmetry breaking terms in $\Gamma_{sb}$. Upon extremizing 
$E(\nu,F]$ for a given value $\nu\in[0,\pi/2]$ the chiral angle 
depends on $\nu$ in a parametrical way. This treatment is to be 
compared with that of the rigid rotator approach where the chiral 
angle is fixed to $F(r,\nu=0)$. These two approaches mainly differ 
by the large distance behavior of the chiral angle: In the RRA the 
chiral angle decays with the pion mass for every flavor orientation. 
On the other hand the SRA exhibits the desired feature that the chiral 
angle decays with the pion mass only for $\nu=0$ while the 
configuration which is maximally rotated into strange direction 
($\nu=\pi/2$) indeed has the kaon mass entering the Yukawa tail.

In the next step the time independence of the flavor rotations 
is waived by substituting the time dependent meson configuration
\be
U(\mbox{\boldmath $r$},t)=A(t)\ 
{\rm exp}\Big(i{\mbox{\boldmath $\tau$}}\cdot
\hat{\mbox{\boldmath $r$}}F(r,\nu)\Big)\ A^\dagger(t)
\label{collrot}
\ee
into the action (\ref{lag}). This allows us to extract a Lagrangian,
which apparently is a function of the time derivative 
of the collective rotation $A$. This derivative is most 
conveniently presented by introducing the angular velocities
$A^\dagger \dot A=(i/2)\sum_{a=1}^8\lambda_a\Omega_a$. The 
canonical quantization introduces the right generators of flavor 
$SU(3)$ via $R_a=-\partial L/\partial \Omega_a$ and leads
to the collective Hamiltonian \cite{Sch92}
\be
H = E(\nu)+\left(\frac{1}{2\alpha^2(\nu)}-
\frac{1}{2\beta^2(\nu)}\right)  {\mbox{\boldmath $J$}}^2
+\frac{1}{2}\left\{\frac{1}{2\beta^2(\nu)}
,C_2\left[ SU(3)\right]\right\}
-\frac{3}{8\beta^2(\nu)}\  
\label{slowham}
\ee
together with the constraint $R_8=\sqrt{3}/2$. This constraint stems 
from the Wess--Zumino term (\ref{WZ}) and guarantees that the 
eigenstates of $H$ possess half--integer spin \cite{Wit83}. In 
eq (\ref{slowham}) $J_i=-R_i$ denotes the spin operator for 
$i=1,2,3$ while $C_2\left[ SU(3)\right]=\sum_{a=1}^8 R_a^2$ refers to 
the quadratic Casimir operator of $SU(3)$. It should be stressed that 
the coefficients in eq (\ref{slowham}), which are functionals of the 
chiral angle, have both an explicit as well as an implicit dependence 
on the strangeness changing angle $\nu$. While the former is due to 
the symmetry breaking part of the action (\ref{sb}) the latter stems 
from the (parametrical) $\nu$ dependence of the chiral angle. The RRA 
corresponds to omitting this implicit dependence and computing the 
coefficients as radial integrals over $F(r,\nu=0)$; the explicit 
dependence in $E(\nu)$ is kept, though. Note that we have 
adopted a symmetric operator ordering to render the Hamiltonian 
Hermitian. As has been shown previously \cite{Sch92} this Hamiltonian 
can be diagonalized exactly. A suitable technique is to express the 
generators $R_a$ as differential operators with respect to the eight 
``Euler angles" parametrizing the rotation matrix $A$. The 
eigenfunctions $\Psi_B(A)=\langle A | B\rangle$ of the collective 
Hamiltonian (\ref{slowham}) are identified as the wave--functions 
corresponding to baryon $B$. These are distorted $SU(3)$ D--functions 
reflecting that in the presence of flavor symmetry breaking the 
resulting baryon eigenstates are no longer pure octet (for $J=1/2$) 
or decouplet (for $J=3/2$) states but rather contain sizable admixtures 
of baryon states with appropriate spin and flavor quantum numbers in 
higher dimensional representations of $SU(3)$ like for example 
${\bf{\overline{10}}}$ or ${\bf 27}$. 

Adjusting the Skyrme parameter to $\epsilon=3.46$ reproduces the 
observed mass differences for the $\frac{1}{2}^+$ and $\frac{3}{2}^+$ 
baryons when the pion and kaon masses 
are taken at their physical values $m_\pi=138{\rm MeV}$ and 
$m_K=495{\rm MeV}$ while the kaon decay constant is chosen to be 
$f_K=118{\rm MeV}$ which is only slightly larger than the experimental 
number $113{\rm MeV}$. Unless otherwise noted we will always adopt 
this set of parameters.

\bigskip
\leftline{\large \it 3. Current Matrix Elements}
\stepcounter{chapter}
\bigskip

Having obtained both the covariant form of the electromagnetic 
current (\ref{curr}) and the baryon wave--functions as eigenfunctions
of the collective Hamiltonian (\ref{slowham}) it is straightforward 
to compute electromagnetic properties of the baryons by evaluating 
the appropriate matrix elements.

The part of the matrix element associated with the out--going photon 
corresponds to taking the Fourier--transform of the electromagnetic 
current $J_\mu$ with respect to the photon momentum $q$ measured in the 
rest frame of the decaying $\frac{3}{2}^+$ baryon. Upon substituting 
the time dependent meson configuration (\ref{collrot}) into the 
covariant expression for the electromagnetic current (\ref{curr}) one 
extracts two operators $\hat E$ and $\hat M$ which have non--vanishing 
transition matrix elements between $\frac{1}{2}^+$ and $\frac{3}{2}^+$ 
baryons. The operators $\hat E$ and $\hat M$ respectively stem from 
the quadrupole part of the time component $J_0$ and the spatial 
component $J_i$:  
\be
\hat E&=&
\int d^3r j_2(qr)\left(\frac{z^2}{r^2}-\frac{1}{3}\right)J_0
\label{defehat} \\
{\hat M}&=&\frac{1}{2}\int d^3r j_1(qr)
\epsilon_{3ij}{\hat r}_i J_j \ .
\label{defmhat}
\ee
Here $j_l(qr)$ denote the spherical Bessel functions corresponding
to orbital angular momentum $l$. Replacing the angular velocities 
$\Omega_a$ in favor of the $SU(3)$ generators $R_a$, $\hat E$ and 
$\hat M$ read
\be
\hat E(q,A,R_a)&=&-\frac{8\pi}{15\alpha^2(\nu)}D_{{\rm em},3}(A)R_3
\int_0^\infty dr\, r^2 j_2(qr)V_0(r,\nu)
\label{ehat}
\ee
and
\be
{\hat M}(q,A,R_a)&=&-\frac{4\pi}{3}\int_0^\infty  dr\, r^3 j_1(qr)
\Bigg[V_1(r,\nu)D_{{\rm em},3}(A)
+V_3(r,\nu)d_{3\alpha\beta}
D_{{\rm em},\alpha}(A)D_{8\beta}(A)
\nonumber\\ &&\hspace{4cm}
-\frac{1}{2}\left\{\frac{1}{\beta^2(\nu)}V_2(r,\nu),
d_{3\alpha\beta}D_{{\rm em},\alpha}(A)R_\beta \right\}\Bigg] \ ,
\label{mhat}
\ee
where $\alpha,\beta=4,\ldots,7$.  Explicit expressions for 
$V_i(r,\nu)$ are listed in the appendix. We have also introduced 
the adjoint representations of the collective rotations 
$D_{ab}=(1/2)$ $\Tr\left(\lambda_a A \lambda_b A^\dagger\right)$.
In particular the subscript ``em" refers to the 
electromagnetic direction: $D_{{\rm em},a}=D_{3a}+D_{8a}/\sqrt3$.
Again we have chosen a symmetric ordering for those terms which 
develop ambiguities when elevating the (classical) angular 
velocities to operators in the space of the collective coordinates.

While $\hat M$ represents the operator whose matrix elements 
directly yield the amplitudes for the $M1$ channel, the operator
$\hat E$ can be associated with the $E2$ channel only after some 
approximations according to Siegert's Theorem \cite{Si37} are 
made\footnote{The matrix elements of $\hat E$ are related to 
the $C_2$ channel.}. When employing the continuity equation 
$\partial_\mu J^\mu$ one has to assume that $j_{l-1}\gg j_{l+1}$ and 
to expand $j_2(qr)\approx (qr)^2/15$. For typical photon momenta
for the radiative hyperon decays ($q\sim 200{\rm MeV}$) this introduces 
errors as little as 10\% since the spatial extension is of the order 
$\langle r^2\rangle \approx 1{\rm fm}^2$ or less. In section 4 we will 
give an estimate of these kinematical corrections. Recently the 
deviation from Siegert's Theorem has been studied in the two flavor 
reduction of the model for the $\Delta$--nucleon transition \cite{Wa96}.
In addition to the before--mentioned kinematical corrections
these authors introduce pion fluctuations off the rotating soliton 
(\ref{collrot}) to consistently satisfy the continuity equation 
$\partial_\mu J^\mu=0$ at subleading order in $1/N_C$. These induced
fields account for shortcomings\footnote{The rotating hedgehog 
(\ref{collrot}) does not represent a solution to the time dependent 
equations of motion, even in the simpler two flavor case.} in the 
collective quantization and brings into the game contributions to 
$E2$ which are not only two orders down in $1/N_C$ compared to those 
parts which are associated with the pure soliton configuration 
(\ref{collrot}) but also have an additional factor $q^2$. Hence one 
is inclined to assume that these corrections are negligibly small. 
However, the numerical studies indicate that this is not quite the 
case. For for the $\Delta$--nucleon transition it has been observed 
\cite{Wa96} that altogether corrections of the order of 30\% may 
arise. They may be split up into those associated with the induced 
field components ($\sim 25$\%) and smaller kinematical ones 
($\sim 5$\%, see below). Unfortunately such an inclusion of induced
fields seems to be unfeasible in the three flavor model with symmetry 
breaking included and in particular in the framework of the SRA. In 
any event, the main purpose of the present investigation is to study 
the inter--relations between various radiative hyperon decays. In this 
regard the structure of the baryon wave--functions $\Psi(A)$ appears 
to be significantly more important than the precise form of the radial 
functions $V_i$. We therefore approximate the $E2$ and $M1$ decay 
widths by
\be
\Gamma_{E2}&=&\frac{675}{8}e^2 q
\left|\langle \Psi_{J=\frac{1}{2}} |{\hat E}(q)|
\Psi_{J=\frac{3}{2}}\rangle\right|^2
\label{GE2}  \\
\Gamma_{M1}&=&18e^2 q
\left|\langle \Psi_{J=\frac{1}{2}} |{\hat M}(q)|
\Psi_{J=\frac{3}{2}}\rangle\right|^2 \ .
\label{GM1}
\ee
These matrix elements are computed as integrals over the collective 
coordinates $A$ using the exact eigenfunctions $\Psi_J(A)$ of the 
collective Hamiltonian (\ref{slowham}). For details we refer to
appendix A of ref \cite{We96}. These analyses also allow us to 
compute the ratio \cite{Wi87}
\be
\frac{E2}{M1}=\frac{5}{4}\
\frac{\langle \Psi_{J=\frac{1}{2}} |{\hat E}(q)|
\Psi_{J=\frac{3}{2}}\rangle}
{\langle \Psi_{J=\frac{1}{2}} |{\hat M}(q)|
\Psi_{J=\frac{3}{2}}\rangle} \ .
\label{e2m1}
\ee

At this point it is interesting to note that the magnetic moments
of the baryons can be obtained from the diagonal matrix elements
of $\hat M$. The computation of the magnetic moments indeed provides
the strongest support for the SRA because it has the distinctive 
feature of reproducing the experimentally observed deviation from 
the $U$--spin symmetric relations between the baryon magnetic 
moments \cite{Sch92}. For example, for the parameters listed above
the ratio $\mu(\Sigma^+)/\mu(p)=0.85$ compares well with the 
experimental number of $0.87$ while assuming $U$--spin symmetry 
gives unity. Similarly, the SRA gives sizable $U$--spin violations 
for the charge radii which in the SRA are predicted to decrease 
with strangeness, similar to the pattern discussed in 
ref \cite{Po90}\footnote{Although these authors study the strong 
interaction radii, which do not need to be the same as the 
electromagnetic ones, they also confirm that ``wherever the comparison 
is possible, the strong interaction radii are very similar to the 
charge radii."}. The comparison with the RRA, wherein $U$--spin 
relations are approximately preserved despite of sizable distortions 
of the collective wave--functions, indicates that only minor 
contributions of the $U$--spin violations stem from configuration 
mixing but rather are due to the influence of symmetry breaking on 
the chiral angle. Similar results are found when the influence of 
symmetry breaking on the extension of the soliton is treated quantum 
mechanically \cite{Sch91}. These inter--relations strongly motivate 
the study of radiative hyperon decays in the SRA. Unfortunately, 
the absolute magnitude of the magnetic moments is underestimated by 
about 30\%. In the two flavor reduction it has been shown that this 
shortcoming is cured when quantum corrections ${\cal O}(1/N_C)$ 
are taken into account \cite{Me96}. Of course, this problem is 
expected to contaminate the predictions for at least the $M1$ 
transition amplitudes such that decay widths are underestimated by 
up to 50\%. However, from quite general arguments it has been found 
that the $1/N_C$ corrections enter the coefficient $V_1$ in eq 
(\ref{mhat}) multiplicatively \cite{Da94}. Hence our conclusions 
concerning the $U$--spin violation and strangeness dependencies of 
the decay widths, which rely on the comparison of various matrix 
elements,  will not be effected crucially.

\bigskip
\leftline{\large \it 4. Numerical Results}
\stepcounter{chapter}
\bigskip

Here we will discuss the numerical results for the electromagnetic
transitions as defined in the preceding section.

In order to estimate the kinematical corrections to Siegert's 
Theorem we also consider a modification of the operator $\hat E$.
According to eq (3.14) of ref \cite{Wa96} we include a Bessel 
function associated with orbital angular momentum $l=3$ into the 
integrand of $\hat E$ (\ref{ehat}):
\be
\hat E^\prime(q,A,R_a)&=&
-\frac{8\pi}{5\alpha^2(\nu)}D_{{\rm em},3}(A)R_3
\int_0^\infty dr\, r^2 
\Big(3j_2(qr)-qrj_3(qr)\Big)V_0(r,\nu) \ .
\label{ehatprime}
\ee
{}From the numerical results displayed in tables \ref{ta_1} and 
\ref{ta_2} we conclude that the kinematical corrections lower 
the $E2$ amplitude by no more than about 5\%. The total 
corrections found in ref \cite{Wa96} by the complicated
treatment of the continuity equation were of the order 30\%. 
As mentioned above the remaining 25\% are due to the induced 
fields\footnote{The effect of these fields associated with the 
modification of the quantization rule 
$R_i=$ $-\partial L/\partial \Omega_i=$
$-\alpha^2\Omega_i+\ldots$ was omitted in ref \cite{Wa96}.} 
at subleading order in $1/N_C$. 

\begin{table}
\caption{\label{ta_1}SRA predictions for the electromagnetic 
decay widths of the $\frac{3}{2}^+$ baryons. Here we consider 
the case $L_9=0$. The data in parentheses correspond to the 
operator $\hat E^\prime(q,A,R_a)$ {\it cf.} eq 
(\protect\ref{ehatprime}).}
~
\newline
\centerline{
\begin{tabular}{|l||c|c|c|} \hline
{\small Transition} &  $\Gamma_{\rm E2}$(eV) & $\Gamma_{\rm M1}$(keV) &
E2/M1(\%) \\
\hline \hline
$\Delta \rightarrow \gamma N           $ & 416.3 (354.6) 
& 326.0 & -2.06 (-1.90) \\
$\Sigma^{*0} \rightarrow\gamma\Lambda  $ & 157.8 (142.2) 
& 163.9 & -1.78 (-1.70) \\
$\Sigma^{*-} \rightarrow\gamma\Sigma^- $ & 1.82  (1.70)
& 1.72  & -1.88 (-1.81)  \\
$\Sigma^{*0} \rightarrow\gamma\Sigma^0 $ & 2.35  (2.25)
& 7.79  & -1.00 (-0.98)  \\
$\Sigma^{*+} \rightarrow\gamma\Sigma^+ $ & 19.5  (18.5)
& 47.5  & -1.17 (-1.14)  \\
$\Xi^{*-}    \rightarrow\gamma\Xi^-    $ & 1.65  (1.54)
& 1.35  & -2.02 (-1.95)  \\
$\Xi^{*0}    \rightarrow\gamma\Xi^0    $ & 29.8  (28.3)
& 64.5  & -1.24 (-1.21)  \\
\hline
\end{tabular}
}
\end{table}
In table \ref{ta_1} the resulting decay widths for the radiative
transitions of the $\frac{3}{2}^+$ baryons are displayed as 
obtained in the SRA. Apparently these widths follow the pattern
\be
\Gamma_{\Delta\to \gamma N}&>&
\Gamma_{\Sigma^{*0}\to\gamma \Lambda}>
\Gamma_{\Xi^{*0}\to\gamma\Xi^0}>
\Gamma_{\Sigma^{*+}\to\gamma\Sigma^+}\gg
\nonumber \\ && \hskip2cm
\Gamma_{\Sigma^{*0}\to\gamma\Sigma^0}\gg
\Gamma_{\Xi^{*-}\to\gamma\Xi^-}\approx
\Gamma_{\Sigma^{*-}\to\gamma\Sigma^-}\approx0 \ .
\label{pattern}
\ee
for both the $M1$ and $E2$ channels. In particular this pattern 
implies that the transition amplitudes for the negatively charged 
$\frac{3}{2}^+$ baryons turn out to be negligibly small.
Thus this $U$--spin relation is maintained in the SRA as well.
\begin{table}
\caption{\label{ta_2}Same as table \protect\ref{ta_1} 
for $L_9=0.0069$.}
~
\newline
\centerline{
\begin{tabular}{|l||c|c|c|} \hline
{\small Transition} &  $\Gamma_{\rm E2}$(eV) & $\Gamma_{\rm M1}$(keV) &
E2/M1(\%) \\
\hline \hline
$\Delta \rightarrow \gamma N           $ & 455.0 (410.7)
& 308.9 & -2.22 (-2.11) \\
$\Sigma^{*0} \rightarrow\gamma\Lambda  $ & 168.7 (158.4)
& 157.4 & -1.89 (-1.83) \\
$\Sigma^{*-} \rightarrow\gamma\Sigma^- $ & 1.90  (1.82)
& 1.67  & -1.95 (-1.91) \\
$\Sigma^{*0} \rightarrow\gamma\Sigma^0 $ & 2.43  (2.37)
& 7.66  & -1.02 (-1.01) \\
$\Sigma^{*+} \rightarrow\gamma\Sigma^+ $ & 20.2  (19.6)
& 46.6  & -1.20 (-1.18) \\
$\Xi^{*-}    \rightarrow\gamma\Xi^-    $ & 1.72  (1.66)
& 1.30  & -2.10 (-2.06)  \\
$\Xi^{*0}    \rightarrow\gamma\Xi^0    $ & 30.9  (30.0)
& 63.2  & -1.28 (-1.26)  \\
\hline
\end{tabular}
}
\end{table}
The non--minimal term (\ref{lagl9}) has only moderate effects 
on the transition matrix elements as can be observed from the 
comparison of tables \ref{ta_1} and \ref{ta_2}. Since the integrands
entering the evaluation of the charge radii and the $E2$ amplitudes
differ only by the additional Bessel function the rise in the electric 
amplitude is expected. However, the total decay widths get reduced 
because the $M1$ amplitudes become smaller when including the 
non--minimal term. Since this decrease effects all channels 
approximately equally we conclude that the $V_1$ contribution in 
eq (\ref{mhat}) dominates the magnetic channel.

As expected from the above discussion on the absolute values 
of the magnetic moments, the total decay width for the 
$\Delta\to\gamma N$ is about 50\% smaller than the empirical 
value $\Gamma_{\Delta\to\gamma N}=610{\rm keV}\ldots730{\rm keV}$ 
given by the PDG \cite{PDG96}. Note that the rescaling motivated by 
the $1/N_C$ analysis \cite{Da94} of the coefficient $V_1$ in eq 
(\ref{mhat}) to account for the absolute value of the proton magnetic 
moment solves this discrepancy as well. Our prediction for the ratio 
$E2/M1$ for the radiative decay of the $\Delta$ resonance is 
surprisingly close to the empirical value $-2.5\pm0.2$ which has 
recently been extracted from the pion--photoproduction experiment 
performed at MAMI \cite{Be95}. However, this agreement should be taken 
with a grain of salt since according to the above discussions we 
expect modifications to the absolute values of both the electric and 
the magnetic transition amplitudes. On the other hand, we think that 
the statement that all $E2/M1$ ratios are small and negative is save 
against the indicated corrections.

Several models \cite{Bu93} as well as dispersion relation analysis 
\cite{Ha96} predict complex values for the ratio $E2/M1$. As the 
subtracted background may be different in these two channels their
ratio is in general not taken at the resonance position. In all 
collective approaches to soliton models, however, the $J=3/2$ baryons 
have zero widths as long as meson fluctuations off the soliton are 
omitted. Hence the resulting ratio $E2/M1$ is real.

To estimate the model dependence we also consider 
the effects of a sixth--order stabilizing term
\be
L_6=-\frac{\epsilon_6^2}{2}B_\mu B^\mu\  
\quad{\rm where}\quad
B_\mu=\frac{1}{24\pi^2}\epsilon_{\mu\nu\rho\sigma}
\Tr\left[\left(U^\dagger D^\nu U\right)
\left(U^\dagger D^\rho U\right) 
\left(U^\dagger D^\sigma U\right)\right] \ .
\label{lsix}
\ee
Note that in the absence of the photon field, $B_\mu$ represents the 
topological current. In the same way in which the Skyrme term mocks 
up the $\rho$ meson exchange the strong interaction part of 
$L_6$ can be interpreted as the exchange of an $\omega$ meson.

\begin{table}
\caption{\label{ta_3}Same as table \protect\ref{ta_1} 
for $f_K=108{\rm MeV}$, $\epsilon=5.61$, 
$\epsilon_6=0.0118{\rm MeV^{-1}}$ and $L_9=0.0069$. With the sixth 
order term (\protect\ref{lsix}) included this set provides a fit 
to the baryon mass differences \protect\cite{Sch92}.}
~
\newline
\centerline{
\begin{tabular}{|l||c|c|c|} \hline
{\small Transition} &  $\Gamma_{\rm E2}$(eV) & $\Gamma_{\rm M1}$(keV) &
E2/M1(\%) \\
\hline \hline
$\Delta \rightarrow \gamma N           $ & 531.1 (478.5)
& 319.3 & -2.35 (-2.24) \\
$\Sigma^{*0} \rightarrow\gamma\Lambda  $ & 210.9 (198.2)
& 185.8 & -1.94 (-1.89) \\
$\Sigma^{*-} \rightarrow\gamma\Sigma^- $ & 1.91  (1.83)
& 1.17  & -2.34 (-2.29) \\
$\Sigma^{*0} \rightarrow\gamma\Sigma^0 $ & 3.83  (3.74)
& 12.3  & -1.02 (-1.01) \\
$\Sigma^{*+} \rightarrow\gamma\Sigma^+ $ & 28.1  (27.3)
& 65.6  & -1.19 (-1.18) \\
$\Xi^{*-}    \rightarrow\gamma\Xi^-    $ & 1.88  (1.80)
& 1.02  & -2.48 (-2.43)  \\
$\Xi^{*0}    \rightarrow\gamma\Xi^0    $ & 44.2  (43.0)
& 94.4  & -1.25 (-1.23)  \\
\hline
\end{tabular}
}
\end{table}
The decay parameters resulting from the inclusion of $L_6$ are shown 
in table \ref{ta_3}. Obviously there are no qualitative changes to the 
ordinary Skyrme model. We find a small increase for most of the decay 
widths. For the $M1$ channel this is expected since in this model the 
predicted proton magnetic moment is $\mu(p)=1.90$ \cite{Sch92} and thus 
a little larger than in the model with only the fourth order stabilizer.
However, the magnetic widths for the $U$--spin forbidden channels
decrease when the sixth order term is included. This corroborates 
our conclusion that these decays remain (at least approximately)
forbidden even when the effects of flavor symmetry breaking are 
incorporated into the soliton configuration.

The electromagnetic transitions of the $\frac{3}{2}^+$ baryons 
have been computed in several other approaches and models. 
In table \ref{ta_4} we display the results from the RRA, {\it i.e.}
we have omitted the implicit dependence on the strangeness 
changing angle $\nu$. Note also that for simplicity we have 
not included the induced kaon fields. Hence these data differ 
slightly from those in ref \cite{Ab96}.
\begin{table}
\caption{\label{ta_4}Same as table \protect\ref{ta_1} 
for the rigid rotator approach:$f_K=113{\rm MeV}$, 
$\epsilon=3.9$, $\epsilon_6=0$ and $L_9=0$.}
~
\newline
\centerline{
\begin{tabular}{|l||c|c|c|} \hline
{\small Transition} &  $\Gamma_{\rm E2}$(eV) & $\Gamma_{\rm M1}$(keV) &
E2/M1(\%) \\
\hline \hline
$\Delta \rightarrow \gamma N           $ & 942.0 (648.4)
& 319.4 & -3.14 (-2.60) \\
$\Sigma^{*0} \rightarrow\gamma\Lambda  $ & 567.9 (403.4)
& 208.9 & -3.01 (-2.54) \\
$\Sigma^{*-} \rightarrow\gamma\Sigma^- $ & 7.47  (6.04)
& 2.27  & -3.31 (-2.98) \\
$\Sigma^{*0} \rightarrow\gamma\Sigma^0 $ & 12.0  (9.71)
& 11.3  & -1.89 (-1.70) \\
$\Sigma^{*+} \rightarrow\gamma\Sigma^+ $ & 93.4  (75.5)
& 67.5  & -2.15 (-1.93) \\
$\Xi^{*-}    \rightarrow\gamma\Xi^-    $ & 12.8  (10.0)
& 3.42  & -3.54 (-3.12)  \\
$\Xi^{*0}    \rightarrow\gamma\Xi^0    $ & 168.3 (131.1)
& 100.2 & -2.37 (-2.09)  \\
\hline
\end{tabular}
}
\end{table}
Apparently the RRA predicts larger $E2$ amplitudes than the SRA while
the changes for the $M1$ amplitudes are not as drastic. Hence the 
$E2/M1$ ratios increase. It should, however, be noted that the 
$M1$ widths involving strange baryons are appreciably smaller 
in the SRA than in the RRA while for the $\Delta\to\gamma N$ 
transition the magnetic widths do not seem to significantly 
depend on the approach. The latter result is due to the fact 
that this channel is dominated by the two flavor portion.
As we favor the SRA over the RRA from their predictions on the 
magnetic moments and the hyperon radii we are lead to the assessment 
that the hyperon radiative decay widths should actually be smaller 
than previously assumed when normalized to the width of the 
$\Delta\to\gamma N$ decay. This is shown in table \ref{ta_5}.
It should be stressed that these numbers will not suffer from 
eventual $1/N_C$ corrections to the $M1$ cannel as discussed at 
the end of section 3.
\begin{table}
\caption{\label{ta_5}Total decay widths normalized to that of the 
$\Delta\to\gamma N$ transition in various models. The results for 
SRA and RRA correspond to tables \protect\ref{ta_2} and
\protect\ref{ta_4}, respectively. The BSA results have been obtained 
with the empirical values for the meson masses and decay constants 
together with $\epsilon=4.25$ and $\epsilon_6=0$. The entry $SU(3)$ 
denotes the predictions from a flavor symmetric formulation of the 
Skyrme model. The data for the quark model (QM) and lattice 
calculation (Lat.) are taken from refs \protect\cite{Da83,Le93}.}
~
\newline
\centerline{
\begin{tabular}{|l||c|c|c|c||c|c|} \hline
{\small Transition} & SRA & RRA & BSA & $SU(3)$ & QM & Lat.  \\
\hline \hline
$\Sigma^{*0} \rightarrow\gamma\Lambda  $ & 0.509 & 0.653 & 0.765 &
3/4 & -- & 0.703 \\
$\Sigma^{*-} \rightarrow\gamma\Sigma^- $ & 0.005 & 0.007 & 0.010 &
0   & 0.007 & 0.006 \\
$\Sigma^{*0} \rightarrow\gamma\Sigma^0 $ & 0.024 & 0.035 & 0.037 &
1/4 & 0.040 & 0.055 \\
$\Sigma^{*+} \rightarrow\gamma\Sigma^+ $ & 0.152 & 0.210 & 0.233 &
1   & 0.233 & 0.303 \\
$\Xi^{*-}    \rightarrow\gamma\Xi^-    $ & 0.004 & 0.011 & 0.039 &
0   & 0.009 & 0.012 \\
$\Xi^{*0}    \rightarrow\gamma\Xi^0    $ & 0.205 & 0.313 & 0.412 &
1   & 0.300 & 0.415 \\
\hline
\end{tabular}
}
\end{table}
Except for the $U$--spin forbidden channels we observe strong 
deviations from the flavor symmetric formulation. Note that the
symmetric formulation also requires to assume a common photon momentum 
for all channels. We see that the inclusion of symmetry breaking 
effects is crucial to maintain the previously established order of the 
decay widths (\ref{pattern}). The deviation from the symmetry relations
is strongest in the SRA where the normalized decay widths apparently 
decrease much more quickly with strangeness than in all other available 
computations. To analyze this behavior it is illuminating to recall 
the predictions of these models on the extensions of the baryons. 
Except of the SRA the radii show only a moderate variation with 
hypercharge \cite{Pa91,Ku90,Le92,Wa95} although the analysis of ref 
\cite{Po90} indicates that the radii should decrease with strangeness. 
In the BSA \cite{Ku90} and the lattice calculation \cite{Le92} the 
electric radius of the $\Sigma^+$ is even slightly larger than that of 
the proton, although in the BSA the $\Xi^-$ radius is predicted 
surprisingly small. The radiative hyperon decays and the radii are 
strongly related as both probe moments of the electromagnetic current. 
It is hence not surprising that in the BSA and the lattice calculation
the decay width $\Gamma_{\Sigma^{*+}\to\gamma\Lambda}$ is significantly
larger than in the RRA and especially in the SRA. From the discussions 
on the magnetic moments and on the hyperon radii it seems fair to 
conclude that pattern of the hyperon decays widths should follow the 
SRA prediction, in particular one would expect a strong decrease 
with strangeness.

\bigskip
\leftline{\large \it 6. Summary and Conclusions}
\stepcounter{chapter}
\bigskip

We have presented a Skyrme model calculation of the widths for 
radiatively decaying $\frac{3}{2}^+$ baryons which will be 
experimentally available soon. Here we have emphasized on discussing 
the flavor symmetry breaking pattern for these widths. From the 
comparison of the model predictions on other observables (magnetic 
moments, radii) we have argued that the slow rotator approach (SRA), 
wherein the stationary equation is solved for each flavor orientation 
of the hedgehog field, is most suitable to address this question. In 
particular this assessment has to be concluded from the observed 
deviations from the $U$--spin relations among the hyperon magnetic 
moments. By including symmetry effects not only in the baryon 
wave--functions but also in the soliton field, which enters the
electromagnetic current, the SRA properly accounts for these 
deviations. This feature made very interesting to examine the 
$U$--spin predictions on the radiative hyperon decays within 
this approach. The $U$--spin predictions state that the processes 
$\Sigma^{*-}\to\gamma\Sigma^-$ and $\Xi^{*-}\to\gamma\Xi^-$ are 
forbidden. Somewhat surprisingly we have found that also in the 
SRA these decay widths are negligibly small. Our main conclusion, 
however, is that the widths should exhibit a stronger dependence 
on strangeness than previously deduced from other model calculations. 
This is meant in the sense that the SRA predicts the decay widths 
to decrease more quickly with strangeness than any other model.
Since the matrix elements entering the computation of the decay 
widths are sensitive to the extension of the hyperons involved
this result is linked to and supported by the empirical pattern 
of the hyperon radii which are also assumed to decrease with 
strangeness.

\bigskip

\appendix
\bigskip
\leftline{\large \it Appendix}
\stepcounter{chapter}
\bigskip

Here we present the explicit expressions for the radial functions
entering the collective operators $\hat E$ and $\hat M$ in 
equations (\ref{ehat}) and (\ref{mhat}), respectively. We also 
include the contributions stemming from the sixth order term 
(\ref{lsix}). To simplify the presentation we introduce the 
abbreviations $s={\rm sin}F$ and $c={\rm cos}F$.
\be
V_0(r,\nu)&=&s^2\left[f_{\pi}^2+
\frac{1}{\epsilon^2}\left(F^{\prime2}+\frac{s^2}{r^2}\right)
+\frac{\epsilon_6^2}{4\pi^2}F^{\prime2}\frac{s^2}{r^2}
+{\rm sin}^2\nu \left(f_K^2-f_\pi^2\right) c \right]
\nonumber \\ &&
-4L_9\left(s^{\prime2}+\frac{s}{r}(rs)^{\prime\prime}
-3\frac{s^2}{r^2}\right) \ ,
\nonumber \\
V_1(r,\nu)&=&\frac{s^2}{r^2}\left[f_\pi^2
+\frac{1}{\epsilon^2}\left(F^{\prime 2}+\frac{s^2}{r^2}\right) 
+\frac{\epsilon_6^2}{4\pi^2}F^{\prime2}\frac{s^2}{r^2}
+{\rm sin}^2\nu \left(f_K^2-f_\pi^2\right) c \right]
\nonumber \\ && 
-\frac{L_9}{r^2}
\left((c^2-s^2)^{\prime\prime}+4\frac{s^2}{r^2}\right) \ ,
\nonumber\\
V_2(r,\nu)&=&\frac{1}{4\pi^2r^2}F^\prime s^2 
\qquad {\rm and} \qquad
V_3(r,\nu)=-\frac{2}{\sqrt{3}}\left(f_K^2-f_\pi^2\right)
         \frac{s^2}{r^2} \ .
\ee
In these expressions a prime indicates the derivative with respect 
to the radial coordinate, {\it i.e.}
$F^\prime=\partial F(r,\nu)/\partial r$. Note, that the 
radial functions $V_i$ not only have an explicit dependence on the
strangeness changing angle $\nu$ but also an implicit one 
via the chiral angle $F(r,\nu)$.

\end{document}